\documentclass[a4paper,fleqn,usenatbib]{mnras}

\usepackage{mathptmx}

\usepackage{longtable}

\usepackage[T1]{fontenc}
\usepackage{ae,aecompl}

\usepackage{graphicx}	
\usepackage{amsmath}	
\usepackage{amssymb}	

\title[Precise CCD positions of Himalia using Gaia DR1]{Precise CCD positions of Himalia using Gaia DR1 in 2015-2016}

\author[H. W. Peng et al.]{
H. W. Peng,$^{1,2,3,4}$
Q. Y. Peng$^{1,4}$\thanks{E-mail: tpengqy@jnu.edu.cn}
and N. Wang$^{1,4}$
\\
$^{1}$Department of Computer Science, Jinan University, Guangzhou 510632, China\\
$^{2}$Yunnan Observatories, Chinese Academy of Sciences, Kunming 650216, China\\
$^{3}$University of Chinese Academy of Sciences, Beijing 100049, China\\
$^{4}$Sino-French Joint Laboratory for Astrometry, Dynamics and Space Science, Jinan University, Guangzhou 510632, China
}

\date{Accepted 2017 January 24. Received 2017 January 19; in original form 2016 October 17}

\pubyear{2016}

\begin{document}
\label{firstpage}
\pagerange{\pageref{firstpage}--\pageref{lastpage}}
\maketitle

\begin{abstract}
In order to obtain high precision CCD positions of Himalia, the sixth Jovian satellite, a total of 598 CCD observations have been obtained during the years 2015-2016. The observations were made by using the 2.4 m and 1 m telescopes administered by Yunnan Observatories over 27 nights. Several factors which would influence the positional precision of Himalia were analyzed, including the reference star catalogue used, the geometric distortion and the phase effect. By taking advantage of its unprecedented positional precision, the recently released catalogue Gaia DR1 was chosen to match reference stars in the CCD frames of both Himalia and open clusters which were observed for deriving the geometric distortion. The latest version of SOFA library was used to calculate the positions of reference stars. The theoretical positions of Himalia were retrieved from the Jet Propulsion Laboratory Horizons System which includes the satellite ephemeris JUP300, while the positions of Jupiter were based on the planetary ephemeris DE431. Our results showed that the means of observed minus computed (O-C) residuals are 0.071 and -0.001 arcsec in right ascension and declination, respectively. Their standard deviations are estimated at about 0.03 arcsec in each direction.

\end{abstract}




\section{Introduction}
\label{Intro}

The irregular satellites in our solar system have been studied for many years. Parts of them may be formed following with their host giant planets. However, some of these objects are widely believed to have been heliocentric asteroids before being captured by a giant planet's gravity (Colombo \& Franklin~\citet{Colombo1971}; Heppenheimer \& Porco~\citet{Heppenheimer1977}; Pollack et al.~\citet{Pollack1979}; Sheppard \& Jewitt~\citet{Sheppard2003}; Agnor \& Hamilton~\citet{Agnor2006}; Nesvorn\'{y} et al.~\citet{Nesvorny2007},~\cite{Nesvorny2014}). This means the irregular satellites may have close relationship with the formation of early solar system. They are smaller and farther, and having highly eccentric and inclined orbits than the regular satellites (Nicholson~\citet{Nicholson2008}; Grav et al.~\citet{Grav2015}). The astrometric observations of irregular satellites are also more difficult to be obtained with high precision. Until now, though many space missions have had close flyby, observations made by ground-based telescopes could still be the primary way to study irregular satellites. The high precision astrometric observations of irregular satellites can also be used for improving the ephemerides of their host planets.

Himalia is the largest member of Jovian irregular satellites (Grav et al.~\citet{Grav2015}). It has been discovered by Perrine at Lick Observatory in 1904 (Perrine~\citet{Perrine1905}). The astrometric observations of Himalia were obtained continuously since then. The precision of positions of Himalia was also improved steadily with the time after larger diameter telescopes and new astrometric data processing techniques were used. Thus two medium size telescopes which are the 2.4 m telescope (Fan et al.~\citet{Fan2015}) and 1 m telescope (Zhou et al.~\citet{Zhou2001}) administered by Yunnan Observatories were used for obtaining our CCD observations. For the data processing, in order to obtain high precision results, several factors were taken into account. Specifically, the reference star catalogue used, the geometric distortion (called GD hereafter) and the phase effect, etc. The latest version of software routines from the IAU-SOFA (International Astronomical Union-Standards of Fundamental Astronomy) collection (Hohenkerk~\citet{Hohenkerk2015}) were used for calculating the topocentric apparent positions of reference stars. Our previously proposed GD solution (Peng et al.~\citet{Peng2012}) was used for deriving the GD patterns. The GD effects of the 2.4 m and 1 m telescopes have been proved in our previous works (Peng et al.~\citet{Peng2012},~\cite{Peng2015}; Zhang et al.~\citet{Zhang2012}; Wang et al.~\citet{Wang2015}; Peng et al.~\citet{Peng2016}). In consideration of the fewer number of reference stars in each CCD frame of Himalia, the high-order plate model can't be applied. In some cases, there are only several reference stars available in the CCD field of view, and a plate model with four constants becomes the most reasonable choice. Under this circumstance, the GD effects should be corrected accurately. Furthermore, the phase effect of Himalia was also analyzed (Lindegren~\citet{Lindegren1977}).

As is well known, the precision of a reference star catalogue is essential to the measurements of irregular satellites. Furthermore, the zonal errors of a star catalogue also have direct influence on the positional precision. The astrometry satellite Gaia (Gaia Collaboration et al.~\citet{GaiaCollaboration2016a}) which is fully funded by the European Space Agency (ESA) has been launched on December 19, 2013. After less than three years, the first catalogue Gaia Data Release 1 (Gaia DR1) (Gaia Collaboration et al.~\citet{GaiaCollaboration2016b}) was published on September 14, 2016. The precise star positions derived by the Gaia could render better predictions with the primary source of error being the ephemerides (de Bruijne~\citet{Bruijne2012}; Gomes-J\'{u}nior et al.~\citet{Gomes2015}). Though the final mission products are waiting to be released in the future, the results of Gaia DR1 still represent a huge improvement in the available fundamental stellar data and practical definition of the optical reference frame (Lindegren et al.~\citet{Lindegren2016}). We believe that the higher positional precision of targets will be achieved.

The contents of this paper are arranged as follows. In Section 2, the CCD observations are described. Section 3 presents the reduction details. In Section 4, results are showed. In Section 5, discussions are made.  Finally, in Section 6, conclusions are drawn.

\section{CCD Observations}
\label{Obs}

A total of 27 nights of CCD observations of Himalia were obtained from two
telescopes administered by Yunnan Observatories during the years 2015-2016. The observational dates were chosen according to the epochs when Jupiter is near its opposition. Specifically, 19 nights of CCD observations were made by the Yunnan Faint Object Spectrograph and Camera (YFOSC) instrument attached to the 2.4 m telescope (longitude E 100$^\circ$1$'$51$''$, latitude N 26$^\circ$42$'$32$''$, height 3193 m above sea level), and 9 nights of CCD observations were made by the 1 m telescope (longitude E 102$^\circ$47$'$18$''$, latitude N 25$^\circ$1$'$46$''$, height 2000 m above sea level). It is noted that observations were obtained from both the 2.4 m and 1 m telescopes on February 12, 2015. The specifications of the two telescopes and corresponding CCD detectors are listed in Table~\ref{Tab1}.

\begin{table}
\centering
\caption{Specifications of the 2.4 m and 1 m telescopes administered by Yunnan Observatories and the corresponding CCD detectors. Column 1 shows the parameters and the following columns list their values for the two telescopes.}
  \label{Tab1}
  \begin{tabular}{rll}
  \hline
  Parameters                           & 2.4 m                        & 1 m            \\
  \hline
  Approximate focal length             & 1920cm                       & 1330cm                   \\
  F-Ratio                              & 8                            & 13                       \\
  Diameter of primary mirror           & 240cm                        & 100cm                    \\
  CCD field of view(effective)         & 9$'\times$9$'$               & 7$'\times$7$'$           \\
  Size of CCD array(effective)         & 1900$\times$1900             & 2048$\times$2048         \\
  Size of pixel                        & 13.5$\mu\times$13.5$\mu$     & 13.5$\mu\times$13.5$\mu$         \\
  Approximate scale factor             & 0.286$''$$/$pixel            & 0.209$''$$/$pixel        \\
  \hline
\end{tabular}
\end{table}

\begin{table}
\centering
\caption{CCD observations of Himalia and calibration fields by using the 2.4 m and 1 m telescopes administered by Yunnan Observatories. Column 1 lists the observational dates. Column 2 shows the open clusters observed. Column 3 and Column 4 list the number of observations for open clusters and Himalia, respectively. The Johnson-I filter was used in all observations. Column 5 shows which telescope was used.}
  \label{Tab2}
  \begin{tabular}{ccccc}
  \hline
  Obs dates        & Calibration fields &               & Himalia       & Tel \\
                   & open clusters      & No.           & No.           &           \\
  \hline\noalign{\smallskip}
  2015-01-31       & NGC1664            & 44            & 21            & 2.4 m     \\
  2015-02-07       & NGC2324            & 44            & 25            & 2.4 m     \\
  2015-02-08       & NGC2324            & 44            & 14            & 2.4 m     \\
  2015-02-09       & NGC2324            & 44            & 18            & 2.4 m     \\
  2015-02-10       & NGC1664            & 44            & 18            & 2.4 m     \\
  2015-02-11       &                    &               & 19            & 2.4 m     \\
  2015-02-12       &                    &               & 17            & 2.4 m     \\
  2015-02-13       &                    &               & 19            & 2.4 m     \\
  2015-02-12       & M35                & 60            & 14            & 1 m       \\
  2015-02-14       &                    &               & 20            & 1 m       \\
  2016-03-01       & M35                & 49            & 36            & 1 m       \\
  2016-03-02       & M35                & 49            & 30            & 1 m       \\
  2016-03-03       & M35                & 49            & 41            & 1 m       \\
  2016-03-04       & M35                & 49            & 51            & 1 m       \\
  2016-03-06       &                    &               & 20            & 2.4 m     \\
  2016-03-08       &                    &               & 18            & 2.4 m     \\
  2016-03-09       &                    &               & 20            & 2.4 m     \\
  2016-03-10       &                    &               & 3             & 2.4 m     \\
  2016-03-11       & NGC6633            & 44            & 23            & 2.4 m     \\
  2016-03-12       &                    &               & 21            & 2.4 m     \\
  2016-03-13       &                    &               & 21            & 2.4 m     \\
  2016-03-14       &                    &               & 24            & 2.4 m     \\
  2016-03-15       &                    &               & 23            & 2.4 m     \\
  2016-03-16       &                    &               & 22            & 2.4 m     \\
  2016-03-17       &                    &               & 20            & 2.4 m     \\
  2016-04-07       &                    &               & 14            & 1 m       \\
  2016-04-09       & M67                & 35            & 11            & 1 m       \\
  2016-04-10       & M67                & 44            & 15            & 1 m       \\
  \hline
  Total            &                    & 599           & 598           &           \\
  \hline
\end{tabular}
\end{table}

A total of 598 CCD observations of Himalia have been obtained, as well as 599 CCD frames of calibration fields which are open clusters. Distributions of the observations with respect to the observational dates are listed in Table~\ref{Tab2}. It can be seen that 366 CCD observations of Himalia were obtained from the 2.4 m telescope, and 232 CCD observations of Himalia were obtained from the 1 m telescope. The exposure time for each CCD frame is ranged from 20 to 120 seconds depending on the telescopes and meteorological conditions. Calibration fields were observed following with Himalia, except for those nights that observations were subjected to rapidly changing weather conditions and limited telescope time.

\begin{figure*}
\centering
\includegraphics[width=1.0\linewidth]{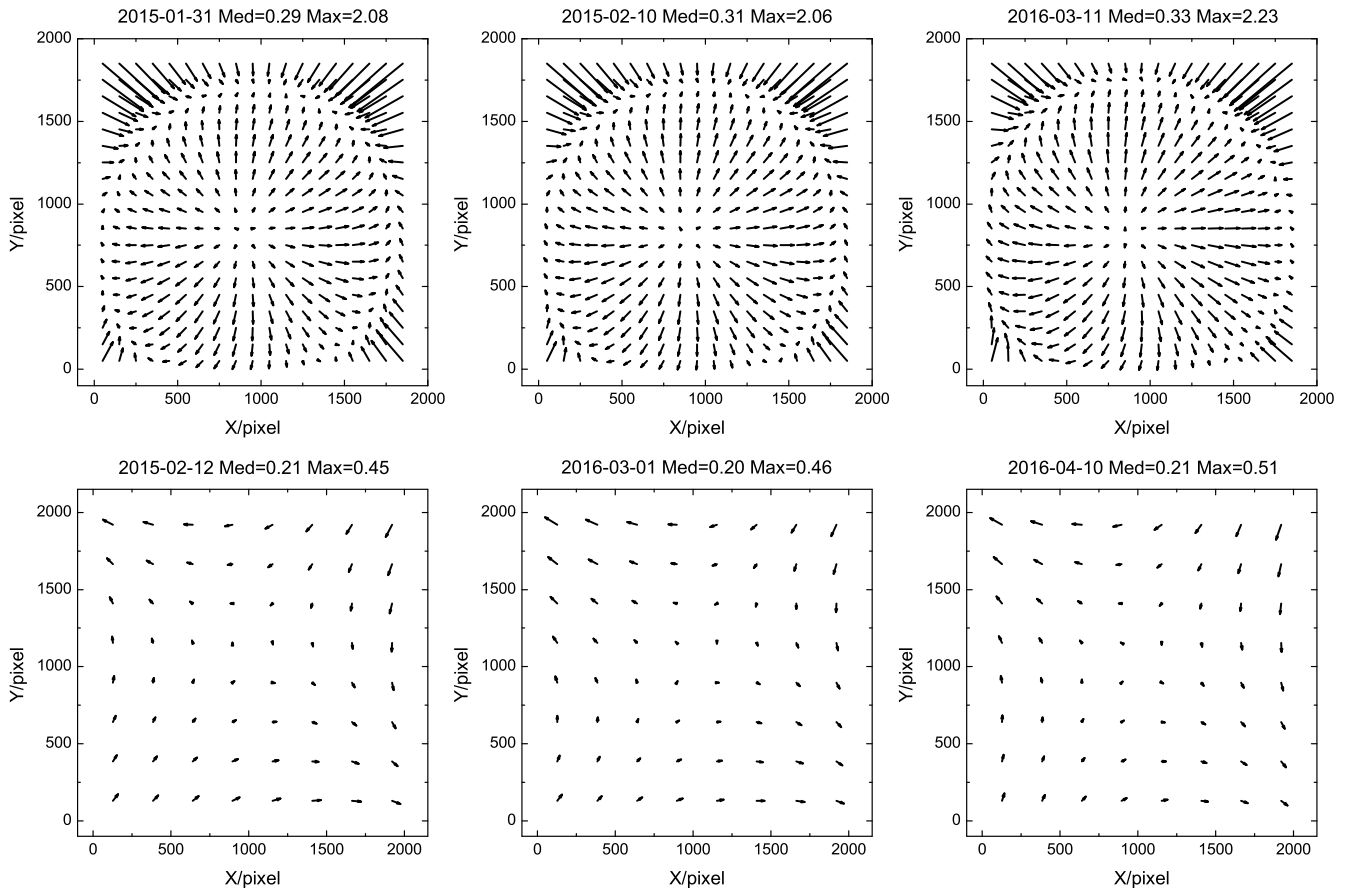}
\caption{GD patterns for the 2.4 m and 1 m telescopes administered by Yunnan Observatories using catalogue Gaia DR1. Only six typical GD patterns are showed. The upper three panels show the GD patterns derived from the CCD observations of the 2.4 m telescope. The lower three panels show the GD patterns derived from the CCD observations of the 1 m telescope. All observations were obtained with the Johnson-I filter. In each panel, the observational date, the median and maximum GD values are listed on the top in units of pixels. A factor of 200 is used to exaggerate the magnitude of each GD vector.}
\label{Fig1}
\end{figure*}

\section{Astrometric reduction}
\label{reduction}

The reduction procedures were carried out according to our previous work (Peng et al.~\cite{Peng2012}). Firstly, all the CCD frames were processed to obtain the pixel positions of both Himalia and reference stars. Secondly, the GD patterns were derived from the CCD frames of open clusters. Thirdly, the GD corrections were applied for both Himalia and reference stars. At last, the topocentric apparent positions of Himalia could be obtained with respect to the reference stars in the same CCD field of view. More description follows.

Before the data reduction, there are several preprocess steps need to be accomplished. As showed in Section~\ref{Obs}, several open clusters were observed for the purpose of deriving GD patterns. These CCD frames together with the CCD frames of Himalia were firstly processed with bias and flat-field corrections. Then the CCD frames obtained from the 2.4 m telescope were clipped into 1900$\times$1900 square pixels because they have ineffective boundaries which have non-exposure pixels. Next, the pixel positions of both Himalia and reference stars were obtained by using the two-dimensional Gaussian fit algorithm. Lastly, a reference star catalogue was chosen to match reference stars in all the CCD frames.

In order to obtain the observational positions of Himalia and also to derive the GD patterns, the topocentric apparent positions of reference stars in all the CCD frames need to be calculated. In this work, the latest version of software routines from the IAU-SOFA collection (Hohenkerk~\citet{Hohenkerk2015}) were used to compute the topocentric apparent positions. However, the catalogue Gaia DR1 does provide parallaxes only for Tycho stars. Thus only these reference stars were applied with both parallaxes and aberration corrections in topocentric correction. The other reference stars were applied with aberration correction, only.

After the positional computations of reference stars in the CCD frames of open clusters were finished, the GD patterns could be derived according to the solution presented in our previous work (Peng et al.~\citet{Peng2012}). For the purpose of deriving GD patterns accurately, the atmospheric refraction effect should be added to the positional computations of reference stars. A standard atmospheric refraction model should be precise enough (Peng et al.~\citet{Peng2012}). According to the study presented in Stone~(\citet{Stone2002}), the filters used and observational zenith distance are important for computing the differential color refraction (DCR). In our observations, the Johnson-I filter was used, and the average of observational zenith distances is about 25 degrees. Under such conditions, the DCR is in some degree negligible according to the study presented in Stone~(\citet{Stone2002}). Thus the DCR for our observations was not taken into account in the refraction model. Fig.~\ref{Fig1} shows six typical GD patterns derived by using the star catalogue Gaia DR1. It can be seen that the GD effects of the 1 m telescope are much smaller than the 2.4 m telescope.

The GD corrections were applied for the pixel positions of both Himalia and reference stars in the same CCD field of view. In practice, GD corrections were applied on each night if the GD pattern is available, otherwise the GD pattern of nearest night is used (Peng et al.~\citet{Peng2016}). More details about the different GD correction schemes are presented in our previous work (Wang et al.~\citet{Wang2015}). After the GD corrections were finished, the topocentric apparent positions of reference stars were calculated. The observational positions of Himalia were computed relative to these reference stars in the same CCD field of view. A plate model with four constants was used for the computations. However, this is accurate only after all the astrometric effects, including the GD effects, are taken into account (Peng et al.~\citet{Peng2012}).

According to the illustration presented in Lindegren~(\citet{Lindegren1977}), the phase effect has a direct influence on positional measurements of planets and natural satellites in our solar system. Phase corrections should be considered and applied for these objects. Though our observational dates were selected near Jupiter's opposition, the phase effect of Himalia is still calculated. As an irregular satellite, Himalia may be far from being a sphere, while this is an implicit assumption in Lindegren~(\citet{Lindegren1977}). Under the assumption that Himalia is a sphere, the phase effect of Himalia was calculated by using equation (14) presented in Lindegren~(\citet{Lindegren1977}). Several typical observational time were selected to compute the phase effects. Our results show that the maximum value of phase effects according to the observational time is as small as $\sim$0.002$''$, and most of the phase effects are less than 0.001$''$. These values are beyond our measurable precision limitation. It means the phase effect of Himalia for our observations is negligible.

\section{Results}
\label{Results}

\begin{figure*}
    \centering
	\includegraphics[width=1.0\linewidth]{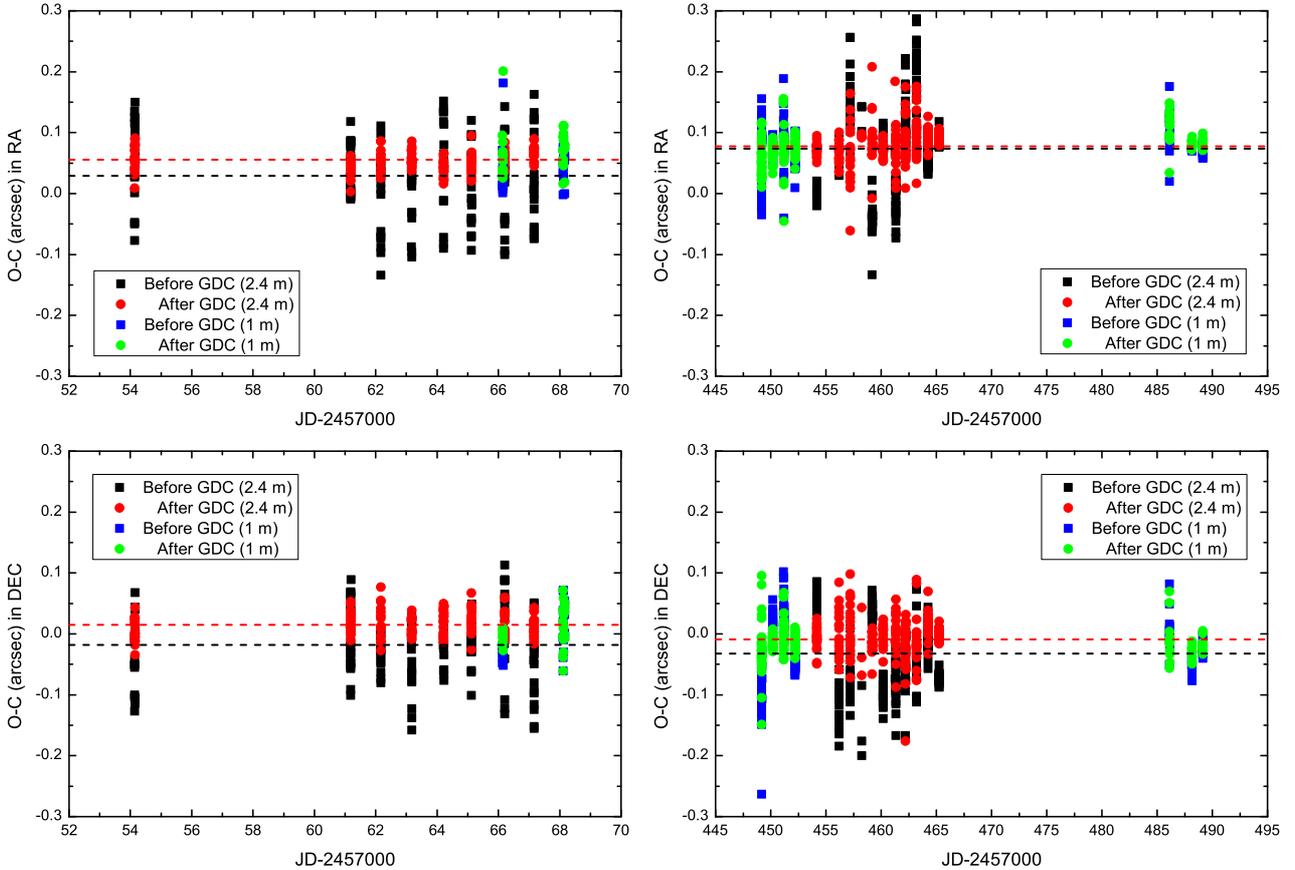}
    \caption{(O-C) residuals of the positions of Himalia in comparison with the ephemeris retrieved from JPL, including the satellite ephemeris JUP300 and planetary ephemeris DE431, with respect to the Julian Dates. The catalogue Gaia DR1 was used for data reduction. The upper two panels show the (O-C) residuals in right ascension in the year 2015 and 2016, respectively. The lower two panels show the (O-C) residuals in declination in the two years. The dark and red points represent the (O-C) residuals for observations obtained from the 2.4 m telescope before and after GD corrections, respectively. The blue and green points represent the (O-C) residuals for observations obtained from the 1 m telescope before and after GD corrections, respectively. In each panel, the dark dash line and red dash line represent the means of (O-C) residuals before and after GD corrections on right ascension and declination for the two years separately.}
    \label{Fig2}
\end{figure*}

\begin{table}
\centering
\caption{Statistics on the (O-C) residuals of positions of Himalia. The catalogue Gaia DR1 was used. Column 1 and column 2 show the observational dates and GD corrections. The following columns list the means of (O-C) residuals and their standard deviations (SDs) in right ascension and declination, respectively. All units are in arcseconds.}
  \label{Tab3}
  \begin{tabular}{rrrrrr}
  \hline
  Obs dates        & GDC       &$\langle$O-C$\rangle$ & SD  &$\langle$O-C$\rangle$ & SD  \\
  and Tel          &           & RA     &        & DEC    &           \\
  \hline
  2015-01-31       & Before     &  0.064 & 0.065  & -0.032 & 0.059     \\
  (2.4 m)          & After      &  0.061 & 0.021  &  0.003 & 0.019     \\
  2015-02-07       & Before     &  0.044 & 0.037  & -0.003 & 0.055     \\
  (2.4 m)          & After      &  0.036 & 0.011  &  0.020 & 0.016     \\
  2015-02-08       & Before     & -0.012 & 0.076  & -0.032 & 0.028     \\
  (2.4 m)          & After      &  0.054 & 0.020  &  0.021 & 0.030     \\
  2015-02-09       & Before     & -0.002 & 0.050  & -0.042 & 0.058     \\
  (2.4 m)          & After      &  0.057 & 0.013  &  0.012 & 0.014     \\
  2015-02-10       & Before     &  0.050 & 0.078  & -0.016 & 0.033     \\
  (2.4 m)          & After      &  0.041 & 0.014  &  0.024 & 0.021     \\
  2015-02-11       & Before     &  0.000 & 0.064  & -0.014 & 0.036     \\
  (2.4 m)          & After      &  0.050 & 0.016  &  0.024 & 0.021     \\
  2015-02-12       & Before     &  0.014 & 0.071  &  0.009 & 0.075     \\
  (2.4 m)          & After      &  0.061 & 0.012  &  0.020 & 0.023     \\
  2015-02-13       & Before     &  0.022 & 0.081  & -0.047 & 0.063     \\
  (2.4 m)          & After      &  0.068 & 0.012  &  0.012 & 0.017     \\
  2015-02-12       & Before     &  0.040 & 0.046  & -0.029 & 0.012     \\
  (1 m)            & After      &  0.063 & 0.045  & -0.003 & 0.011     \\
  2015-02-14       & Before     &  0.054 & 0.027  &  0.014 & 0.035     \\
  (1 m)            & After      &  0.075 & 0.028  &  0.013 & 0.034     \\
  2016-03-01       & Before     &  0.051 & 0.053  & -0.092 & 0.045     \\
  (1 m)            & After      &  0.054 & 0.026  & -0.024 & 0.043     \\
  2016-03-02       & Before     &  0.071 & 0.014  &  0.023 & 0.011     \\
  (1 m)            & After      &  0.067 & 0.014  & -0.013 & 0.010     \\
  2016-03-03       & Before     &  0.082 & 0.034  &  0.030 & 0.026     \\
  (1 m)            & After      &  0.074 & 0.031  &  0.002 & 0.026     \\
  2016-03-04       & Before     &  0.070 & 0.017  & -0.038 & 0.013     \\
  (1 m)            & After      &  0.077 & 0.012  & -0.014 & 0.011     \\
  2016-03-06       & Before     &  0.000 & 0.010  &  0.056 & 0.019     \\
  (2.4 m)          & After      &  0.074 & 0.010  & -0.006 & 0.019     \\
  2016-03-08       & Before     &  0.064 & 0.019  & -0.116 & 0.039     \\
  (2.4 m)          & After      &  0.066 & 0.018  &  0.006 & 0.039     \\
  2016-03-09       & Before     &  0.155 & 0.049  & -0.058 & 0.042     \\
  (2.4 m)          & After      &  0.061 & 0.051  &  0.007 & 0.044     \\
  2016-03-10       & Before     &  0.129 & 0.023  & -0.154 & 0.061     \\
  (2.4 m)          & After      &  0.095 & 0.003  & -0.021 & 0.058     \\
  2015-03-11       & Before     & -0.036 & 0.038  &  0.036 & 0.019     \\
  (2.4 m)          & After      &  0.089 & 0.038  & -0.015 & 0.016     \\
  2016-03-12       & Before     &  0.090 & 0.012  & -0.097 & 0.018     \\
  (2.4 m)          & After      &  0.072 & 0.010  & -0.003 & 0.018     \\
  2016-03-13       & Before     & -0.006 & 0.033  & -0.078 & 0.035     \\
  (2.4 m)          & After      &  0.078 & 0.037  &  0.000 & 0.034     \\
  2016-03-14       & Before     &  0.128 & 0.042  & -0.041 & 0.038     \\
  (2.4 m)          & After      &  0.085 & 0.037  & -0.024 & 0.043     \\
  2016-03-15       & Before     &  0.204 & 0.040  & -0.042 & 0.042     \\
  (2.4 m)          & After      &  0.100 & 0.040  & -0.014 & 0.039     \\
  2016-03-16       & Before     &  0.056 & 0.015  & -0.010 & 0.021     \\
  (2.4 m)          & After      &  0.091 & 0.015  &  0.013 & 0.021     \\
  2016-03-17       & Before     &  0.092 & 0.009  & -0.072 & 0.009     \\
  (2.4 m)          & After      &  0.054 & 0.008  &  0.000 & 0.010     \\
  2016-04-07       & Before     &  0.106 & 0.037  &  0.004 & 0.032    \\
  (1 m)            & After      &  0.111 & 0.029  & -0.009 & 0.034     \\
  2016-04-09       & Before     &  0.078 & 0.006  & -0.055 & 0.012     \\
  (1 m)            & After      &  0.083 & 0.006  & -0.030 & 0.012     \\
  2016-04-10       & Before     &  0.072 & 0.008  & -0.022 & 0.011     \\
  (1 m)            & After      &  0.084 & 0.008  & -0.018 & 0.010     \\
  Total            & Before     &  0.060 & 0.066  & -0.028 & 0.056     \\
                   & After      &  0.071 & 0.029  & -0.001 & 0.030     \\
  \hline
\end{tabular}
\end{table}

In this work, the reference star catalogue Gaia DR1 (Gaia Collaboration et al.~\citet{GaiaCollaboration2016b}) was used in the astrometric data reduction. The observed positions of Himalia were compared to the ephemeris retrieved from Jet Propulsion Laboratory (JPL) Horizons ephemeris service (Giorgini et al.~\citet{Giorgini1996}) which includes the satellite ephemeris JUP300 (Jacobson~\citet{Jacobson2013}) and planetary ephemeris DE431 (Folkner et al.~\citet{Folkner2014}). Fig.~\ref{Fig2} shows the (O-C) residuals of positions of Himalia with respect to the Julian Dates. Table~\ref{Tab3} lists the statistics on the (O-C) residuals for Himalia before and after GD corrections. It can be seen that the internal agreement or precision for an individual night has been significantly improved after GD corrections for the 2.4 m telescope, but the results slightly improved for the 1 m telescope. This is due to the GD effects associated with the 1 m telescope being quite smaller than the 2.4 m telescope. The means of (O-C) residuals for all data sets after GD corrections are 0.071$''$ and -0.001$''$ in right ascension and declination, respectively. Their standard deviations are estimated at about 0.03$''$ in each direction. We can see that this precision is improved by a factor of two than the results before GD corrections.

\begin{figure*}
    \centering
	\includegraphics[width=1.0\linewidth]{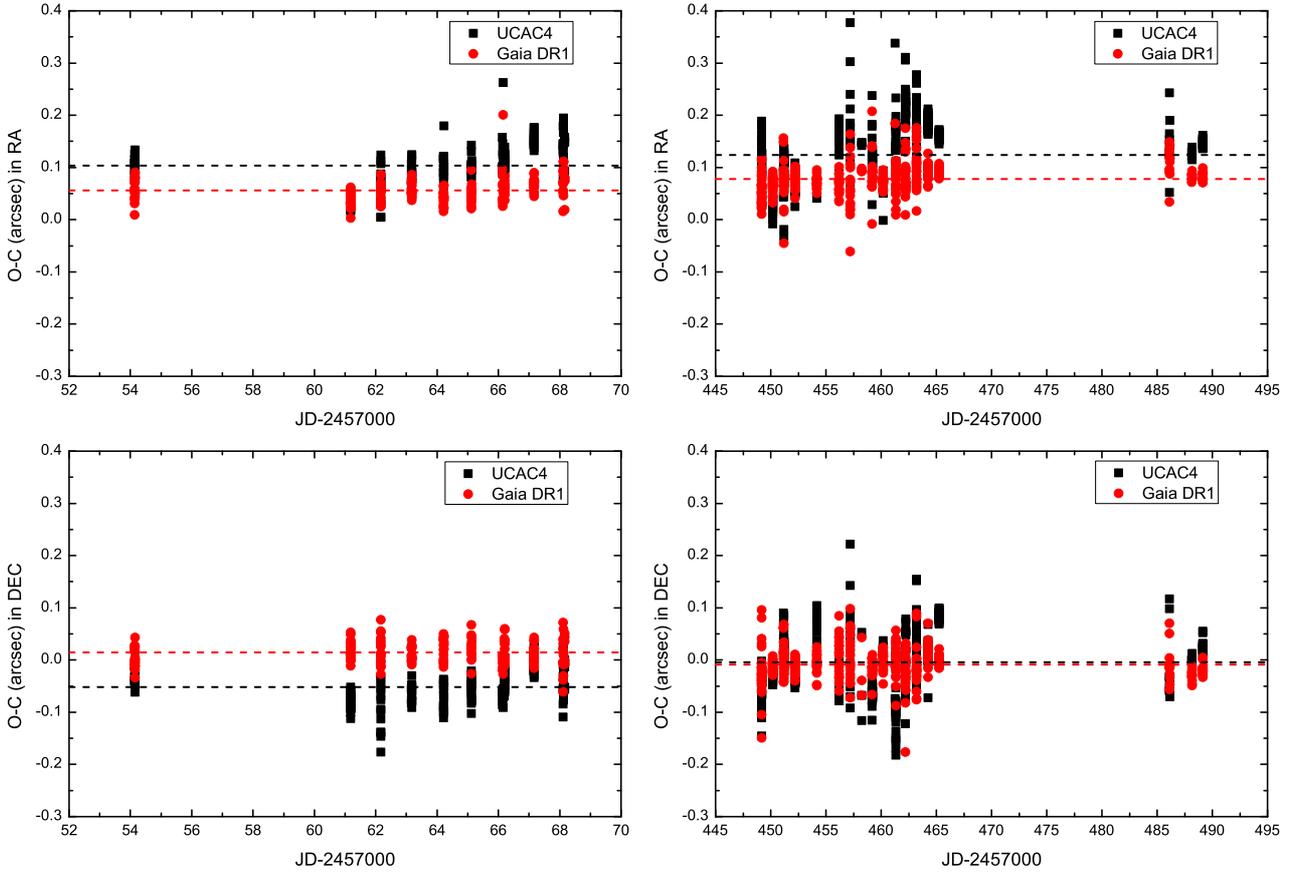}
    \caption{(O-C) residuals of the positions of Himalia in comparison with the ephemeris retrieved from JPL which includes the satellite ephemeris JUP300 and planetary ephemeris DE431, with respect to the Julian Dates. The catalogue UCAC4 and Gaia DR1 were both used. The upper two panels show the (O-C) residuals in right ascension in the year 2015 and 2016, respectively. The lower two panels show the (O-C) residuals in declination in the two years. The dark points represent the (O-C) residuals after GD corrections for UCAC4 and the red ones represent the (O-C) residuals after GD corrections for Gaia DR1. In each panel, the dark dash line and red dash line represent the means of (O-C) residuals after GD corrections on right ascension and declination for the two years separately.}
    \label{Fig3}
\end{figure*}

In order to analyze the effects on positional precision made by the reference star catalogue, the catalogue UCAC4 (Zacharias et al.~\citet{Zacharias2013}) was also used for astrometric reduction. Fig.~\ref{Fig3} shows the (O-C) residuals of positions of Himalia using UCAC4 and Gaia DR1. The same ephemeris which was retrieved from JPL was used. It can be seen larger systematic errors appearing in the (O-C) residuals of Himalia by using UCAC4 than the results by using Gaia DR1, both in right ascension and declination. This may be mainly caused by the existence of zonal errors in the reference star catalogue UCAC4. However, the systematic errors by using Gaia DR1 are significantly reduced. Table~\ref{Tab4} shows the statistics on the (O-C) residuals for Himalia for UCAC4 and Gaia DR1, respectively. It can be seen that the positional precision using Gaia DR1 has been improved by a factor of two than the results using UCAC4. This is mainly resulting from the unprecedent precision of reference star catalogue Gaia DR1. Our results give proof that the tremendous potential in improving astrometric precision of CCD observations of irregular satellites by using Gaia DR1.

\begin{table}
\centering
\caption{Statistics on the (O-C) residuals of positions of Himalia for Gaia DR1 and UCAC4. Column 1 shows the number of CCD observations. Column 2 shows the catalogue used. The following columns list the means of (O-C) residuals and their standard deviations in right ascension and declination, respectively. The ephemerides were retrieved from JPL which include the satellite ephemeris JUP300 and planetary ephemeris DE431. All units are in arcseconds.}
\label{Tab4}
  \begin{tabular}{cccccc}
  \hline
  N   &  Catalogue      & $\langle$O-C$\rangle$ & SD    & $\langle$O-C$\rangle$ & SD    \\
      &                 & RA           &       & DEC          &       \\
  \hline
  598 & UCAC4           &  0.118  & 0.058 & -0.019  & 0.055 \\
      & Gaia DR1        &  0.071  & 0.029 & -0.001  & 0.030 \\
  \hline
\end{tabular}
\end{table}

\section{Discussions}

\begin{figure}
    \centering \includegraphics[width=1.0\linewidth]{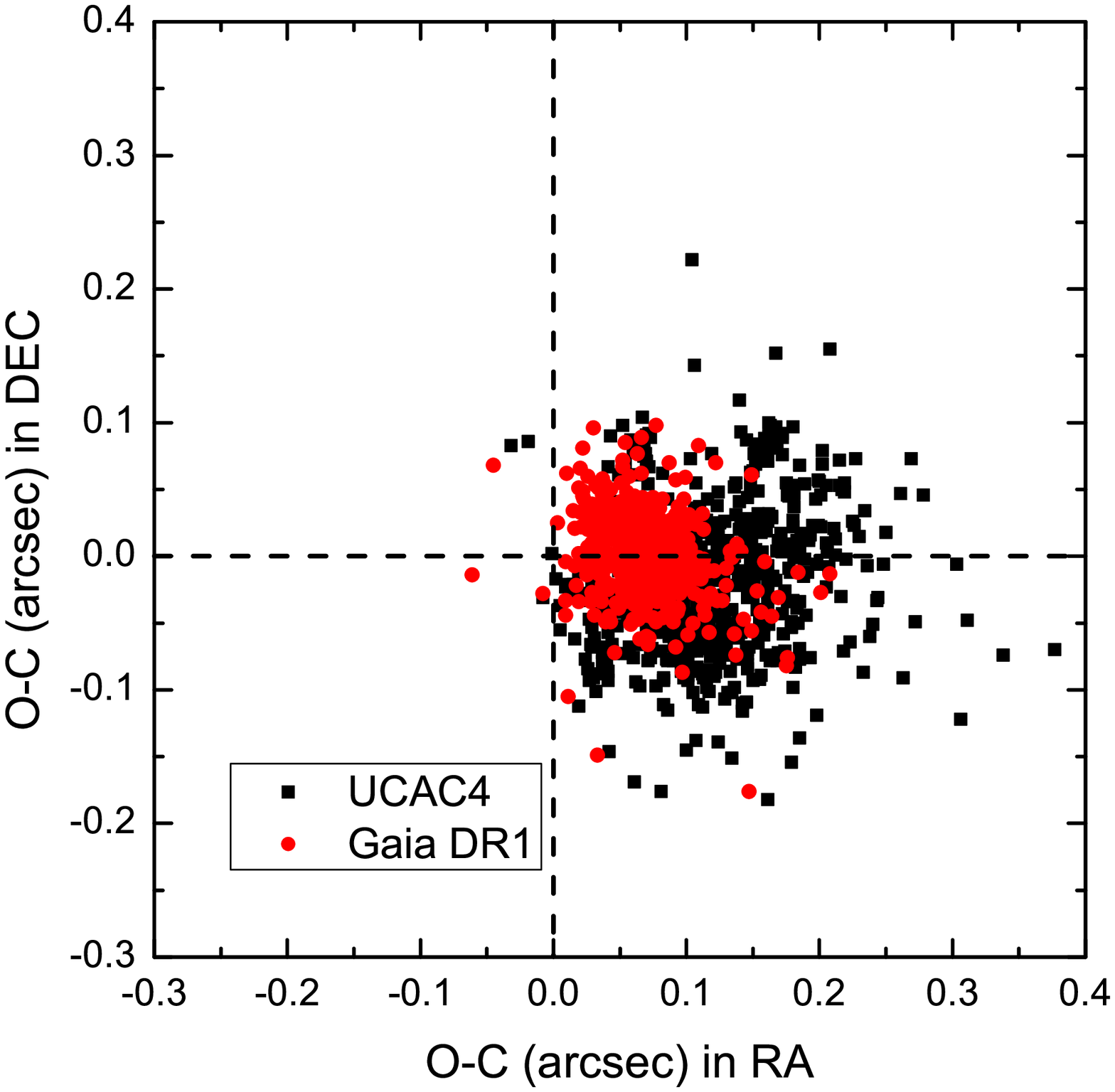}
    \caption{(O-C) residuals of Himalia in declination with respect to right ascension. The dark points represent the (O-C) residuals after GD corrections for catalogue UCAC4 and the red ones represent the (O-C) residuals after GD corrections for catalogue Gaia DR1. The ephemeris was retrieved from JPL which includes the satellite ephemeris JUP300 and planetary ephemeris DE431. The two dark dash lines represent zero lines in each direction.}
    \label{Fig4}
\end{figure}

In order to analyze the dispersion of (O-C) residuals of Himalia, the (O-C) residuals of Himalia by using both catalogue Gaia DR1 and UCAC4 are drawn together. Fig.~\ref{Fig4} shows the details. It can been seen that the dispersion of (O-C) residuals for Gaia DR1 is much more compact than UCAC4. The systematic errors appearing in the (O-C) residuals of Himalia by using Gaia DR1 are significantly reduced. This is because the zonal errors existing in catalogue Gaia DR1 are quite subtle. We can also see from Fig.~\ref{Fig2} that the dispersion of (O-C) residuals for Himalia after GD corrections in the year 2016 is somewhat worse than the results in the year 2015. This is mainly due to the seeing of observations in 2016 is worse than the seeing in 2015.

\begin{figure*}
    \centering
	\includegraphics[width=1.0\linewidth]{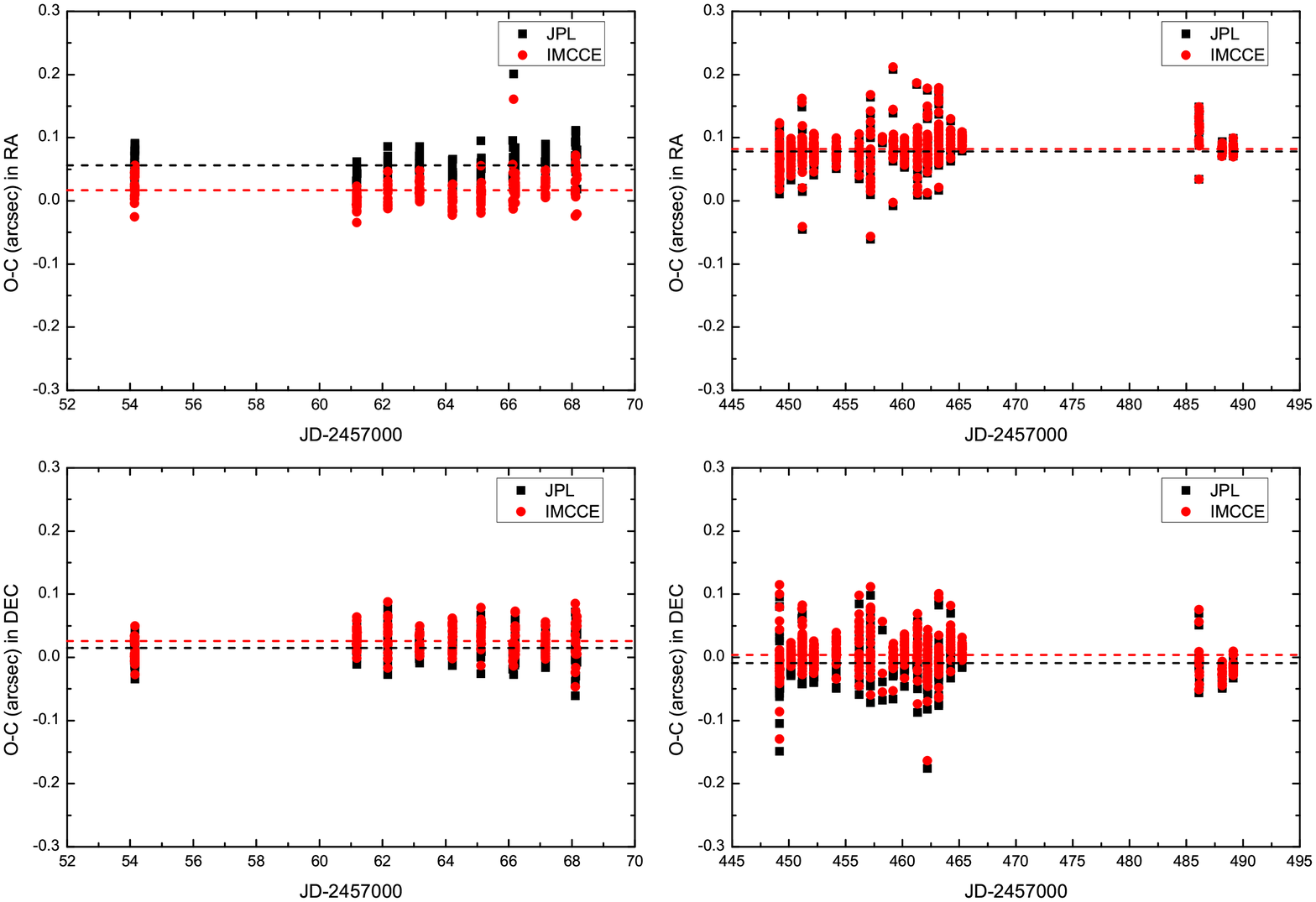}
    \caption{(O-C) residuals of the positions of Himalia in comparison with the two different ephemerides. The upper two panels show the (O-C) residuals in right ascension in the year 2015 and 2016, respectively. The lower two panels show the (O-C) residuals in declination in the two years. The dark points represent the (O-C) residuals after GD corrections for the ephemeris retrieved from IMCCE which includes the satellite ephemeris by Emelyanov~(\citet{Emelyanov2005}) and planetary ephemeris DE431. The red points represent the (O-C) residuals after GD corrections for the ephemeris retrieved from JPL which includes the satellite ephemeris JUP300 and planetary ephemeris DE431. In each panel, the dark dash line and red dash line represent the means of (O-C) residuals after GD corrections on right ascension and declination for the two years separately.}
    \label{Fig5}
\end{figure*}

For comparison, the ephemerides retrieved from Institute de M\'{e}chanique C\'{e}leste et de Calcul des \'{E}ph\'{e}m\'{e}rides (IMCCE) were also obtained, including the satellite ephemeris by Emelyanov~(\citet{Emelyanov2005}) and planetary ephemeris DE431. Fig.~\ref{Fig5} shows the (O-C) residuals of topocentric astrometric positions of Himalia in comparison with the two different ephemerides. Table~\ref{Tab5} shows the statistics on the (O-C) residuals for Himalia after GD corrections for the two ephemerides. The means of (O-C) residuals for ephemeris retrieved from IMCCE are 0.062$''$ and 0.011$''$ in right ascension and declination, respectively. Their corresponding standard deviations are 0.040$''$ and 0.030$''$. The means of (O-C) residuals for all data sets between the two ephemerides are nearly equal. We can also see from Table~\ref{Tab5} that their standard deviations between the two ephemerides for the two years separately are in good agreement. However, a difference for the means of (O-C) residuals in right ascension between the two ephemerides in 2015 could be found. In consideration of the same planetary ephemeris DE431 was used, this difference should be from the different satellite ephemerides used.

\begin{table}
\centering
\caption{Statistics on the (O-C) residuals for Himalia in comparison with the two different ephemerides. Column 1 shows the number of CCD observations. Column 2 shows the ephemeris used. The following columns list the means of (O-C) residuals and their standard deviations in right ascension and declination, respectively. The ephemerides retrieved from IMCCE and JPL are Emelyanov~(\citet{Emelyanov2005})/DE431 and JUP300/DE431, respectively. All units are in arcseconds.}
\label{Tab5}
  \begin{tabular}{rrrrrrr}
  \hline
  Year  & N   &  Ephemeris      & $\langle$O-C$\rangle$ & SD    & $\langle$O-C$\rangle$ & SD    \\
        &     &                 & RA      &       & DEC     &       \\
  \hline
  2015  & 185 & IMCCE           &  0.017  & 0.023 &  0.026  & 0.023 \\
        &     & JPL             &  0.056  & 0.023 &  0.015  & 0.022 \\
  2016  & 413 & IMCCE           &  0.082  & 0.029 &  0.004  & 0.030 \\
        &     & JPL             &  0.078  & 0.029 & -0.009  & 0.030 \\
  \hline
  Total & 598 & IMCCE           &  0.062  & 0.040 &  0.011  & 0.030 \\
        &     & JPL             &  0.071  & 0.029 & -0.001  & 0.030 \\
  \hline
\end{tabular}
\end{table}

To compare our CCD observations with previous ones, some major observational statistics of Himalia are listed in Table~\ref{Tab6}. The data are retrieved from the Minor Planet Center (MPC). The ephemeris used for our observations was retrieved from JPL which includes the satellite ephemeris JUP300 and planetary ephemeris DE431. Table~\ref{Tab6} shows the statistics on the (O-C) residuals. The positions of Himalia are observed topocentric astrometric positions. It can be seen that our positional precision has significant improvements.

\begin{table}
\centering
\caption{Comparisons made with other observations retrieved from the MPC. Column 1 shows the IAU code of observatory. Column 2 lists the number of CCD observations. The following columns show the mean (O-C) residuals and its standard deviation (SD) in right ascension and declination, respectively. All positions of Himalia are observed topocentric astrometric positions. All units are in arcseconds.}
\label{Tab6}
  \begin{tabular}{rrrrrrr}
  \hline
  IAU            &  No. & $\langle$O-C$\rangle$ & SD    & $\langle$O-C$\rangle$ & SD    & Time(yr)  \\
  code           &      & RA      &       & DEC     &       &           \\
  \hline
  689            & 277    & -0.013  & 0.163 & -0.013  & 0.160 & 2000-2007  \\
  415            & 24     & -0.008  & 0.106 &  0.065  & 0.088 & 2008       \\
  809            & 23     & -0.051  & 0.092 &  0.014  & 0.045 & 2007-2009  \\
  511            & 357    & -0.021  & 0.049 & -0.008  & 0.061 & 1997-2008  \\
  874            & 56     & -0.039  & 0.112 & -0.026  & 0.070 & 1992-2014  \\
  874            & 238    & -0.077  & 0.175 & -0.009  & 0.034 & 1992-2014  \\
  874            & 560    &  0.001  & 0.069 & -0.018  & 0.053 & 1992-2014  \\
  Our            & 598    &  0.071  & 0.029 & -0.001  & 0.030 & 2015-2016  \\
  \hline
\end{tabular}
\end{table}

Table~\ref{Tab7} lists the extract example of the observed topocentric astrometric positions of Himalia. The positions are listed with respect to Julian Date (UTC). RA which expressed in hours, minutes and seconds are the observed topocentric astrometric positions of Himalia in right ascension. DEC are the observed topocentric astrometric positions of Himalia in declination, expressed in degrees, arcminutes and arcseconds.

\begin{table}
\centering
\caption{Extract example of the observed topocentric astrometric positions of Himalia. These positions are comparable to the astrometric positions of stars at the same pixel locations with Himalia. According to Urban \& Seidelmann (\citet{Urban2013}), only proper motion and annual parallaxes are taken into account in the astrometric place computation of stars. Thus the aberration is ignored. The observed topocentric astrometric positions of Himalia in this paper can also be directly compared to the positions in its astrometric ephemeris. Column 1 shows the exposure middle time of each CCD observation in form of Julian Date (UTC). RA and DEC are the observed topocentric astrometric positions in right ascension and declination, respectively.}
\label{Tab7}
  \begin{tabular}{ccc}
  \hline
  Date              &  RA             & DEC                   \\
   JD               &  h m s          & $^\circ$ $'$ $''$     \\
  \hline
  2457 054.134 398    & 09 22 28.8650     & 16 45 24.733      \\
  2457 054.136 921    & 09 22 28.8028     & 16 45 25.122      \\
  2457 054.138 322    & 09 22 28.7675     & 16 45 25.338      \\
  \ldots\ldots        & \ldots\ldots      & \ldots\ldots      \\
  2457 489.135 787    & 11 03 12.1353     & 07 16 08.834      \\
  2457 489.136 609    & 11 03 12.1150     & 07 16 08.979      \\
  2457 489.137 488    & 11 03 12.0936     & 07 16 09.134      \\
  \hline
\end{tabular}
\end{table}

\section{Conclusions}

In this paper, a total of 598 CCD observations obtained from the 2.4 m and 1 m telescopes administered by Yunnan Observatories were processed. Several factors were analyzed, including the reference star catalogue used, the geometric distortion and the phase effect. The reference star catalogue Gaia DR1 and UCAC4 were both used for astrometric reduction and have been made with comparisons. Positional precision of Himalia has been significantly improved after GD corrections. The systematic errors existing in the results by using UCAC4 are significantly reduced after Gaia DR1 was used. Comparisons between the two different ephemerides which retrieved from JPL and IMCCE have been made. Our results show that the means of (O-C) residuals of Himalia are 0.071$''$ and -0.001$''$ by using ephemeris retrieved from JPL in right ascension and declination, respectively. Their standard deviations are estimated at about 0.03$''$ in each direction. This positional precision of Himalia is significantly improved by taking advantage of the unprecedent precision of star catalogue Gaia DR1 in comparison with previous works.

\section*{Acknowledgements}

This research work is financially supported by the National Natural Science Foundation of China (grant nos. U1431227,11273014). We acknowledge the support of the staff of the Lijiang 2.4 m telescope. Funding for the telescope has been provided by CAS and the People's Government of Yunnan Province. We also acknowledge the support of the staff of the 1 m telescope at Yunnan Observatories. This work has made use of data from the European Space Agency (ESA) mission {\it Gaia} (\url{http://www.cosmos.esa.int/gaia}), processed by the {\it Gaia} Data Processing and Analysis Consortium (DPAC, \url{http://www.cosmos.esa.int/web/gaia/dpac/consortium}). Funding for the DPAC has been provided by national institutions, in particular the institutions participating in the {\it Gaia} Multilateral Agreement.









\bsp	
\label{lastpage}
\end{document}